# Bringing The Apple to the Moon.

J. A. M. Pereira*

*Departamento de Física - Universidade Federal do Estado do Rio de Janeiro - UNIRIO.*

*Av. Pasteur 458 - Urca 22290-240, Rio de Janeiro, Brasil*

**PACS: 04.08.-y, 91.10.-v, 96.20.Jz, 01.55.+b, 01.40.-d, 01.65.+g**

*key words: Gravitation, Mechanics, Physics Education, History of Physics*

**Abstract**

*Newtonian gravity can be regarded as a hypothetic - deductive system where the inverse square law is the starting point from which gravitational phenomena are deduced. This operational form of presenting gravity endorses problem solving and seems to be predominant in the teaching practice. In contrast, regarding phenomena as a source for the development of the theory is also possible, of course, and can be advantageous to scientific education since it deals with model conception and construction. This article intends to introduce undergraduates to Newtonian gravity using its empirical basis, i.e. the free fall and the planetary motion, to deduce the universal law of gravitation. Additionally, an elemental analysis of the theory´s structure is presented. It also steps into the modern interpretation of gravitational phenomena i.e. Einstein´s general relativity, including a discussion on the instantaneous action at a distance in this context. This didactic presentation of the theory of gravity differs from the standard approach usually found in common textbooks. It is designed to reach a threefold equality, similar to those applied in the method of separation of variables in partial differential equations, where $G$ is treated as a separation constant. By doing so, the universality of the gravitation constant emerges as a conclusion rather than a statement. It is also meant to create a perception on how imagination can be helpful in discovering Physical laws.*

## 1. Introduction

The famous Newton´s universal law of gravitation appeared in the year 1687 in the Principia. It brings a way to calculate an attraction force between two masses $m$ and $M$ apart a distance $r$ [1]. Its simplest modern formulation reads

$$F_G = G \frac{mM}{r^2} \qquad (1)$$

*corresponding author: jpereira.cefeteq@gmail.com

where $r$ is the distance between the centres of the two bodies, $F_G$ is the magnitude of the gravitational force and $G$ is the gravitational constant ($G = 6.67 \times 10^{-11}$ N m$^2$/kg$^2$) [2]. From the educational point of view, equation (1) is usually taught like a sudden inspiration of Isaac Newton when he was at the Woolsthorpe Manor, in Lincolnshire, and noticed the free fall of an apple [3,4]. The universal law of gravitation is then taken as a fundamental postulate from which one deduces gravitational phenomena. This form of presenting gravity seems to be predominant in teaching practice since in most university physics courses [5,6,7]; and also in advanced mechanics [4,8,9]; equation (1) is presented "ready to be used". This cut off is actually desirable when the point is to solve problems related to classical mechanics, the explanation of Kepler´s three laws being one important example. However, the reverse path is pedagogically helpful: regarding phenomena as a source for the development of theories can be instructive to scientific education since it deals with model conception and construction [10].

Historically, it is worth mentioning that the notorious apple was never mentioned in the Principia and that a mathematical expression like equation (1) was not written by Newton. The universal law of gravitation appears in book III – Propositions VII and VIII of the Principia - where equation (1) is described in words [1]. In 1803, Siméon D. Poisson, acknowledging the works of Pierre S. Laplace on celestial mechanics [11], may have been the first to write algebraically that the force of gravity is jointly proportional to the two masses, $mM$, and inversely proportional to the square distance, $1/r^2$, in his Treatise on Mechanics [12]. In this text, the character $f$ is used to name what is known today as the gravitational constant ($G$ in equation 1) and interpreted as an intensity factor defined as the attraction force between two unit masses apart a unit distance. However, the value of the gravitational constant was unknown at that time albeit the 1798 Henry Cavendish research on the torsion balance was already published [13]. As a matter of fact, the Cavendish measurements aimed the determination of the mean mass density of the Earth and, although torsion balances can be used to measure the gravitational constant, the concept of a universal constant (such as $G$) was yet to come. An estimate for the gravitational constant, still referred by $f$ as in Poisson´s notation, based on the Earth´s density found by Cavendish, appeared in 1873 in a scientific communication by Cornu and Baille. [14]. Several measurements of the gravitational constant surfaced during the nineteenth century turn [15] and the problem of the gravitational force as we know today, i.e. equation (1), appears in the works of John H.

Poynting concerning Earth´s science [16,17]. In Poynting´s 1892 paper [17], the gravitational constant, already referred by the symbol G, is given as $\frac{6.66}{10^8}$ in CGS units and this was the value known to Albert Einstein by the time general relativity was published in 1915 [18]. There are earlier references to the ´big G´ in the literature [19].

Considering a contemporary analysis of the gravitational force law, another important characteristic of equation (1) is that the masses appearing on its numerator are referred as gravitational mass which, in modern science, stands for the ability to interact through a gravitational field. This is to be distinguished from the inertial mass, appearing in Newton´s second law, which is a measure of inertia i.e. the resistance that an object presents to a change on its state of movement. It should be stressed that Newton did not distinguish these two mass concepts. For him, inertia and gravity are properties of mass which, in turn, is defined as quantity of matter [20]. The distinction between inertial and gravitational mass is important because these two concepts are logically independent and preserve causality in the application of Newton´s second law to problems involving gravity. The roots of these different mass concepts are related to the equivalence principle and traces back to times before Galileo Galilei [21]. A counterintuitive consequence of this principle is that all bodies fall with the same acceleration irrespectively to their masses. This was one of many important findings Galileo reported in the Dialogues; a book that broke much of the Aristotelic tradition in Physics [22]. In modern terms, the fact that free fall does not depend on the falling object´s mass consists of what is called the 'weak equivalence principle'. The gravitational mass concept bridges gravity towards a field theory and is crucial in Einstein´s general relativity. Actually, Einstein used the terms inertial mass and gravitational mass as suggested at the end of one of his 1907 paper [23]. At Einstein´s time the most successful quantitative testing of the equivalence principle was done by Eötvös using a torsion balance [24]. Satellite experiments are now available, and show agreement between the gravitational and inertial mass values within one part in $10^{18}$ [25]. There is a further sub categorization of gravitational mass, introduced by Bondi, into active gravitational mass, that originates the gravitational field, and passive gravitational mass, that concerns the intensity of the gravitational attraction [26].

One aim of this article is to present the universal law of gravitation from first principles, using its empirical basis and letting the reasoning guide the mathematical procedures. As in the usual approach, a few basic characteristics of the Earth - Moon system are used in the analysis proposed here. Nonetheless, the proposed discussion fully integrates the universal law of gravitation with Newton´s three laws of movement in a way that makes the universality of the constant $G$ self-evident being a conclusion rather than a statement. The proposed didactic sequence is formed by a line of thought designed to reach a threefold equality, similar to those applied in the method of separation of variables in partial differential equations [27], where $G$ plays the role of the separation constant. The material in the didactic sequence, presented in the next section, is written in view of the student´s (or freshman´s) reading so the arguments outlined below form one feasible route to the law of gravity, keeping the way as simple as possible. This is done in a step by step approach so the reader can follow the reasoning in small and precise amounts. Only very basic Physics concepts are required: the kinematics of constant acceleration (either linear or centripetal), Newton´s three laws of movement, Kepler´s planetary laws [28], Galileo´s free fall and the concept of centre of mass. It is important to mention that the sequence presented here (section 2) does not follow Newton´s own line of reasoning and that the modern terminology for mass is used aiming a brief discussion of today´s interpretation of the theory of gravity i.e., Einstein´s general theory of relativity, done at section 3. The sequence is presented with numbered arguments that are referred to each other and a short summary is presented lately for closure.

**2. The didactic sequence**

The intended argument is as follows:

**I** - The daily experience shows that any suspended object, an apple for instance, will fall to the ground when left free. According to Newton´s first law - inertia - a force must be acting on the object, otherwise it would keep motionless in its original position after being left. Remarkably, the fall movement does not depend on the specific place where the experience is done. As long as the distance between the falling object and the Earth´s surface is small when compared to the Earth´s radius, its type of movement is always uniformly accelerated.

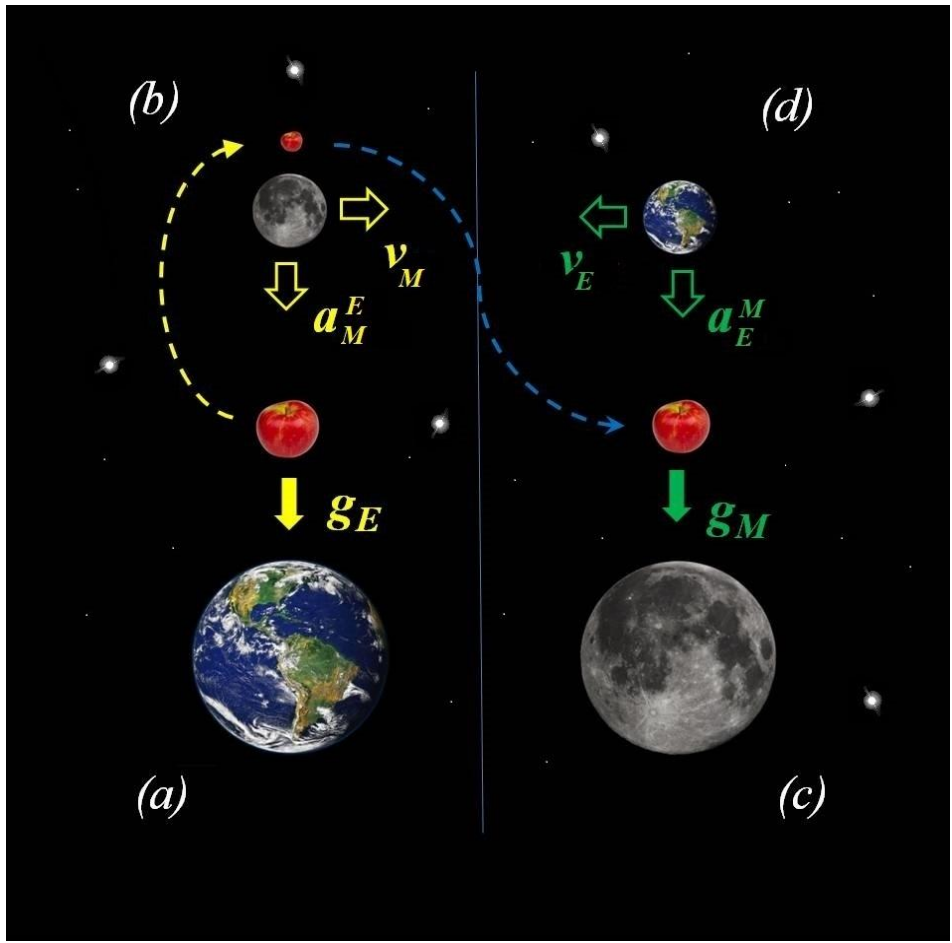

*Figure 1 - The schematic view of the system treated in the didactic section. The left part of the figure, (a) and (b), refers to arguments **I** to **V** while the right one, (c) and (d), refers to arguments **V** to **VII**. (a) A falling apple near the surface of the Earth with acceleration $g_E$. (b) The Moon with velocity $v_M$ and acceleration $a_M^E$. The orbiting apple moves along with the Moon sharing its orbital dynamics (see argument **V** and footnote 2). (c) The same apple falling on the surface of the Moon with acceleration $g_M$. (d) The Earth with velocity $v_E$ and acceleration $a_E^M$. All the accelerations and velocities are to be computed in the centre of mass of the respective system (see argument **XI** and footnote 3).*

**II** - When left free, the movement of the apple is then driven by Newton´s second law – dynamics - which is written as:

$$F_{EA} = mg_E \qquad (2)$$

where $F_{EA}$ is the force of gravity that the Earth exerts on the apple and $m$ is the inertial mass of the apple. It should be emphasised that $g_E$ in equation (2) stands for acceleration, not gravitational field. It is also important to mention that the acceleration $g_E$ is always directed to the centre of the Earth (figure 1a).

**III** - If the apple falls on the Earth surface, why the Moon does not do the same? A response to this question may be given considering that the Moon presents itself to us always with the same size. From this fact, one concludes that the Moon, in spite of having a non-vanishing velocity, as is clear by its periodic movement relative to the "fixed" stars, keeps its distance to the Earth´s surface. A type of movement compatible with this observation is the uniform circular motion. Going back to argument **I**, there must be a force producing the centripetal acceleration for this to happen, otherwise, the Moon would have a rectilinear uniform motion flying away from Earth and vanishing from the sky view. The opening question of this argument can then be answered by saying that the Moon actually keeps falling on Earth but, because of Moon´s velocity (value and direction) it always miss the Earth surface since, given that the Earth is round, its ground is never there for the Moon to hit it (figure 1b). The orbiting Moon moves parallel to the Earth´s ground while the falling apple does not, hence their completely different types of movement. Yet, similarly to the case of the apple´s free fall, the Moon´s acceleration is directed to the centre of the Earth. Hence, one can deduce that the Earth´s gravity produces a centripetal acceleration on all the objects around it.

**IV** - Having the forces bringing the apple to the Earth´s ground and the one keeping the Moon orbiting the Earth the same direction, as seen in arguments **II** and **III**, one can apply Newton´s second law to the movement of the Moon in the same way it was done for the apple of argument **I**. Actually, this is part of an inductive argument made by Newton in the Principia (Book III - Proposition IV – Theorem IV) when he declares that "...*the force by which the Moon is retained in its orbit is that very same force we commonly call gravity* ...". Letting $F_{EM}$ be the force the Earth exerts on the Moon, $M_M$ the inertial mass of the Moon and $a_M^E$ the acceleration of the Moon due to the Earth´s gravity (figure 1b) one has that[1]:

$$F_{EM} = M_M a_M^E \qquad (3)$$

**V** - The experience of being on the surface of a massive giant spheroid is just repeated if one could go to the Moon bringing the apple of argument **I** and execute a free fall experiment with it up there (figures 1b and 1c). The similarity between the two situations compels us to consider that to an observer at the Moon´s surface, the apple movement will be qualitatively the same as the one near the Earth´s surface that is, uniformly accelerated (figure 1c). If $g_M$ is the acceleration due to Moon´s gravity near its surface one has that:

---

[1] *In advanced mechanics, the acceleration of a two dimensional movement in polar coordinates is given by $a = \ddot{r} + r\dot{\theta}^2$[8]. Following this notation, one has that $g_E = \ddot{r}$, since for the apple on Earth $\dot{\theta} = 0$ and $a_M^E = r\dot{\theta}^2$ if the radial movement of the Moon is considered to be negligible.*

$$F_{MA} = mg_M \tag{4}$$

as was done in argument **II** for the terrestrial case (equation 2). In equation (4), $F_{MA}$ is the force that the Moon exerts on the apple, $m$ is the inertial mass of the apple and, as in equation (2), the symbol $g_M$ stands for acceleration not gravitational field. This corresponds to Newton´s second law to the fall of the apple near the surface of the Moon[2] and is a logical extrapolation of the validity of Newton´s laws outside Earth. It is important to stress that care must be taken when applying Newton´s second law to non inertial frames such as the Moon´s surface. For this particular application, equation (4) gives the correct result (see footnote 2).

**VI** - Similarly, one can use argument **III** and **IV** the other way around to write the force, $F_{ME}$, the Moon does on Earth

$$F_{ME} = M_E a_E^M \tag{5}$$

where $M_E$ is the inertial mass of the Earth and $a_E^M$ is the acceleration of Earth due to the Moon´s gravity (figure 1d).

**VII** - Under the assumption that the forces expressed in equations (3) and (5) obey Newton´s third law – action - reaction – they should have the same magnitude and opposite directions ($\vec{F}_{ME} = -\vec{F}_{ME}$) and, consequently, the ratio of their absolute value is unitary. It follows from equations (3) and (5) that

$$\frac{|\vec{a}_M^E|}{|\vec{a}_E^M|} = \frac{M_E}{M_M} \tag{6}$$

It should be noticed that equation (6) sets the centre of mass of the Earth-Moon system to be the frame in which the values of the accelerations $a_M^E$ and $a_E^M$ are to be computed since the sum of the mass - acceleration products of each body is zero in this case: $M_M \vec{a}_M^E + M_E \vec{a}_E^M = 0$. It is also important to mention that if equations (3) and (5) are indeed an action - reaction pair, the Earth-Moon relative movement can be decoupled from the motion of their centre of mass meaning they can be studied separately. In a sense, the assumption that the forces represented by equations (3) and (5) are equal in magnitude can be seen as a verifiable assumption (the instantaneous action at a distance hypothesis) and it is instructive to analyze the consequences of having $|\vec{F}_{ME}| \neq |\vec{F}_{ME}|$. This is done in appendix A where it is shown that, in this case, the internal forces would affect the motion of the system´s centre of mass leading to the violation of conservation laws [20].

---

[2]*From the point of view of someone on the Earth, an observer at the Moon´s surface is in an accelerated frame where inertial (fictitious) forces apply. Therefore, strictly speaking, three forces have to be considered in the description of the apple´s fall for an observer on the Moon: the apple´s weight in the lunar surface ($mg_M$), the centripetal force responsible for the apple´s orbit around the Earth ($mr\dot{\theta}_{orbiting\ apple}^2$), since $\dot{\theta} \neq 0$ for the orbiting apple, and the fictitious force, which corresponds to the inertial mass of the apple times minus the Moon´s centripetal acceleration ($-mr\dot{\theta}_{Moon}^2$). It occurs that, since the orbiting apple and the Moon are in the same orbit, the apple moves relatively to the Earth in the same way the Moon´s ground does. Consequently, one has that $\dot{\theta}_{orbiting\ apple} = \dot{\theta}_{Moon}$ so the fictitious force cancels the centripetal force on the orbiting apple. As a result, equation (4) is the net resulting force on the apple for an observer on the Moon´s surface. (see also section 11.1 of reference [6] for a more detailed discussion on inertial forces in this context).*

**VIII** - In general, the accelerations $\vec{a}_M^E$ and $\vec{a}_E^M$ can be writen in terms of the centre of mass and of the relative accelerations, $\vec{A}$ and $\vec{a}$ respectively, as (see equation A2):

$$\begin{cases} \vec{a}_M^E = \vec{A} + \frac{1}{1+(M_M/M_E)} \vec{a} \\ \vec{a}_E^M = \vec{A} - \frac{(M_M/M_E)}{1+(M_M/M_E)} \vec{a} \end{cases} \quad (7)$$

Following the previous argument, one has that $\vec{A} = 0$ in the centre of mass frame and, since the Moon is much lighter than the Earth ($M_M << M_E$)[3], equation (7) gives $\vec{a}_M^E \sim \vec{a}$ and $\vec{a}_E^M \sim -(M_M/M_E)\vec{a}$. That means the Earth-Moon relative acceleration is truly a very good approximation for the value of Moon´s acceleration, $a_M^E$, defined in equation (3). The study of the Moon´s orbit dynamics can then be done replacing the Moon´s acceleration with respect to the centre of mass by the Earth-Moon relative acceleration with the advantage that the latter is easily evaluated without the precise knowledge of the masses $M_E$ and $M_M$ (see also argument **XI**).

**IX** - The value of the Earth - Moon relative acceleration, $a$, can be found from the radius, $r$, and sidereal period, $T$, of the Moon´s orbit[3] as:

---

[3]*The center of mass of the Earth-Moon system is located inside the Earth at position $R = \frac{M_M}{M_E+M_M} r$ from the center of the Earth where $r$ is the Earth - Moon distance ( $r \sim 3.84 \times 10^8 m$ ). One has the ratio $(M_M/M_E) \sim (1 / 81)$ and therefore R<<r. In addition, the sidereal period of the Moon (27,3 days) gives the angular velocity of the rotation movement by $\omega = 2\pi/T$ and determines the values of $v_M = \omega (r - R) \sim \omega r \sim 1,02 \ km/s$, $v_E = \omega R \sim 12 \ m/s$, $a_M^E = \omega^2 (r - R) \sim \omega^2 r \sim 2.7 \times 10^{-3}$ and $a_E^M = \omega^2 R \sim 3.3 \times 10^{-5}$ (see figure 1).*

$$a = \frac{v^2}{r} = \frac{4\pi^2 r}{T^2} \qquad (8)$$

where $v = 2\pi r/T$ is Moon´s orbital velocity. Equation (8) corresponds to the centripetal acceleration mentioned in argument **III**. A second relationship between $r$ and $T$ is available from Kepler´s third law:

$$T^2 = k\, r^3 \qquad (9)$$

where the pre-factor $k$ is a proportionality constant. By merging equations (8) and (9) one finds that the centripetal acceleration in (8) should vary as the inverse square law:

$$a = \frac{4\pi^2}{k}\frac{1}{r^2} \qquad (10)$$

The appendix B shows that equations (1), (9) and (10) hold for the more general case of elliptical orbits [29].

**X** - The application of equation (10) to the problem of the apple falling on Earth can be done by viewing the distance $r$, in equation (10), as a variable and noting that the accelerations $a(= a_M^E)$ and $g_E$ are both directed to the centre of the Earth (figures 1$a$ and 1$b$). The centre of mass of the apple-Earth system is virtually the centre of the Earth and, therefore, the relevant distance to evaluate the apple´s acceleration near the Earth surface is the radius of the Earth, $R_E$ (see also argument **XI** below). So, one has just to substitute $r$ by $R_E$ and $a$ by $g_E$ in equation (10), resulting in

$$g_E = \frac{4\pi^2}{k}\frac{1}{R_E^2} \qquad (11)$$

In view of the approximation mentioned in argument **VIII**, one can simply replace the $a$ in equation 10 by $a_M^E$ and compare the ratio $[a_M^E/g_E]$ to the ratio $[r/R_E]$. One finds that:

$$\frac{a_M^E}{g_E} = \left[\frac{R_E}{r}\right]^2 \qquad (12)$$

letting us write

$$a_M^E = \frac{R_E^2}{r^2} g_E \qquad (13)$$

**XI** - The numerical values involved in equation (12) were key to convince Newton of the inverse square law of equation (12) as he realized that the calculation of the ratio $[r/R_E]$ is ~ 60 while the ratio $[g_E/a_M^E]$ gives ~ 3600 which is $60^2$ as equation (12) suggests [1,29,30]. As said before, the approximation regarding the accelerations in argument **VIII** ($\vec{a}_M^E \sim \vec{a}$) permit us to take the relative acceleration in place of the acceleration of the Moon with respect to the centre of mass for the present purpose. If this was not the case, it would be much more difficult to reach the inverse square law since the passage from equation (10) to equation (12) would necessarily involve precise knowledge of the masses $M_E$ and $M_M$ values which were not available at Newton´s time. In a sense, it is fortunate that the Moon to Earth mass ratio is very small so the relative acceleration can be used by someone on Earth to establish the inverse square law without greater difficulties.

By itself, equation (13) is a manifestation the principle of universality since it links the acceleration of a terrestrial object, $g_E$, to the acceleration of a celestial body, $a_M^E$. The remaining arguments push universality further. It should be said that Newton argued that the universality of the Physical laws is a fundamental principle, as stated in his four rules of philosophizing [1,30,31].

**XII** - The ratio expressed in equation (12) ought to be valid in the lunar situation given the similarity between the two cases (figure 1$d$). In other words, the ratio of the acceleration of the Earth, $a_E^M$, to the acceleration of gravity on the surface of the Moon, $g_M$, should be proportional to the square ratio between the radius of the Moon, $R_M$, and its distance to Earth, $r$. Thus, one has that

$$\frac{a_E^M}{g_M} = \left[\frac{R_M}{r}\right]^2 \tag{14}$$

similarly to equation (12). It follows that

$$a_E^M = \frac{R_M^2}{r^2} g_M \tag{15}$$

It should be stressed that, contrary to argument **X**, the relative acceleration cannot be used to replace $a_E^M$ (as was done in equation (12) for $a_M^E$) because $a$ is not a good approximation for $a_E^M$ since the ratio $M_M/M_E$ has to be taken in consideration as seen in arguments **VIII** and **XI**. Nevertheless, equation (14) give correct results as long as $a_E^M$ is taken as the Earth´s acceleration due to Moon´s gravity as seen in the centre of mass of the Earth-Moon system.

**XIII -** If equations (13) and (15) are used in equation (6) one can write, after a little algebra:

$$\frac{g_E R_E^2}{M_E} = \frac{g_M R_M^2}{M_M} \tag{16}$$

Equation (16) expresses that the value obtained by the combination of $g$, the acceleration of a falling object near the surface of a planet, $R$, the radius of the planet and $M$, the inertial mass of the planet, should not depend on where we are (Earth or Moon in this case). Therefore, it must have a universal value which can be called $G$. Consequently one has that,

$$G = \frac{g_E R_E^2}{M_E} = \frac{g_M R_M^2}{M_M} \tag{17}$$

or in general

$$G = \frac{gR^2}{M} \tag{18}$$

Equation (17) is central in the development of this sequence and ties the Newtonian universality principle into a mathematical expression.

**XIV** - Being a universal constant, $G$ controls, in every case, the scale of the relationship between $g$, $M$, and $R$ whatever their value are (known or not known). Then, from equation (18), one can deduce the value of the acceleration of gravity on the surface of any planet if one is provided with values of the universal constant of gravitation, $G$, of the planet´s inertial mass, $M$, and the distance from the surface to the centre of the planet (or its radius), $R$:

$$g = G\frac{M}{R^2} \tag{19}$$

**XV** - Equation (19) can be used together with equations (2) or (4) to evaluate the force on an apple due to gravity on the surface of the Earth or on the surface of the Moon as follows:

$$F_{EA} = G\frac{mM_E}{R_E^2} \tag{20a}$$

$$F_{MA} = G\frac{mM_M}{R_M^2} \tag{20b}$$

Finally, the attraction force between the Earth and the Moon is found using equation (3) together with (13) and the definition of $g$ presented in equation (19) applied to Earth.

$$F_{EM} = M_M a_M^E \tag{21a}$$

$$a_M^E = \frac{g_E R_E^2}{r^2} \tag{21b}$$

$$g_E = G\frac{M_E}{R_E^2} \tag{21c}$$

By bringing together (21$a$, $b$ and $c$) one has:

$$F_{EM} = G\frac{M_M M_E}{r^2} \tag{22}$$

**XVI** - Equation (22) can be generalized so any two objects of inertial masses $m$ and $M$ apart a distance $r$, will sense a force whose magnitude is given by

$$F = G\frac{Mm}{r^2} \qquad (23)$$

The physical quantities presented so far are all related to the kinematic effects that gravity produces on the movement of falling or orbiting bodies. In other words, an object suffering a centripetal force given by equation (23) will bend its trajectory producing the observed orbit. However, the above arguments do not mention what causes the bodies to attract one another and, as a matter of fact, Newton himself wrote in the general scholium at the end of the principia that "... *we have explained the phenomena of the heavens and of our sea by the power of gravity, but have not yet assigned the cause of this power* …" [1]. Hence, equation (23) is not fully consistent with dynamics since it cannot be assigned to the cause of movement. One has to consider that Newton´s second law is a causal relation so the cause of movement (the force in the left hand side) must contain elements that are independent of the effects (the mass acceleration product on the right hand side).

**XVII** - Although equation (23) has the same mathematical form as equation (1), none of the above arguments mentioned the gravitational mass, hence the conceptual distinction between gravitational mass, $m^G$, and inertial mass, $m^I$, which are concepts from modern Physics, must now come into play. In other words, what was just shown is that equation (23) can be deduced from the universality of the three laws of movement, including the instantaneous action at a distance hypothesis, plus Kepler´s third law. Therefore, the universal law of gravitation would be merely a corollary of Newtonian mechanics if it was not for the weak equivalence principle ($m^I = m^G$) that makes equation (23) conceptually different from equation (1). The equivalence between gravitational and inertial masses bridges equation (23) into equation (1) substituting inertial mass in equation (23) by gravitational mass in equation (1). This final step imprints causality to the universal law of gravitation and also brings another meaning for equation (19), in the sense that $g$ can be reinterpreted as a gravitational field if $M = M^G$ (the case $m^I \neq m^G$ is analyzed in appendix A). Strictly speaking, the dynamical description of gravitational phenomena in Newtonian mechanics should be written equating (1) to (23) ($F_G = F$):

$$G \frac{M^G m^G}{r^2} = m^I a \tag{24}$$

**3. Discussion**

Within the Newtonian view of nature, the universal law of gravitation is synthesized as a force law like equation (1) because of the law of inertia as seen in argument **I**. In the above analysis, the inverse square law results from the combination of Kepler´s third law and the formulation of centripetal acceleration for circular motion as established in argument **IX** (see appendix B for elliptical orbits). The proportionality of the force of gravity to the joint product of masses is related to the combination of the law of dynamics and the concept of the centre of mass (along with the law of action-reaction) as seen in arguments **IV**, **VII** (equation 6) and **VIII**.

A subtle suggestion in the assumption made in arguments **VII** and **XIII**, that shapes the mathematical form of equation (23), is the existence of instantaneous action at a distance since there is no contact between the two bodies in question. However, Newton felt unease with this feature of the theory, as can be seen in the following quote from his letters to Bentley: *"That one body may act upon another at a distance through a vacuum, without the mediation of anything else, by and through which their action and force may be conveyed from one to another, is to me so great an absurdity that, I believe, no man who has in philosophic matters a competent faculty of thinking could ever fall into it…"* [32]. It is the phenomenological range covered by Newtonian gravity that lead us to conclude for its validity (within the classical Physics limits).

Finally, the universality of the gravitational constant, $G$, expressed physically in argument **XI** and mathematically in equation (17), brings closure to the universal law of gravity. It is what makes possible to apply the three Newtonian laws of movement to astronomical objects like comets, planetary systems and so on. The discovery of the outer planets of the Solar System, for instance, took place thinking of the disturbed orbits of known planets to be caused by the reaction force of an unseen body [33]. The missing object could be located at a precise point of space according to the predictions of Newtonian gravity. Conversely, for terrestrial applications, one could say that one significant characteristic of Cavendish paper is that his experiment showed that the force of gravity occurs for non-celestial bodies providing experimental proof that all matter, celestial or non-celestial, is subjected to gravitational attraction [15]. The measurement of the gravitational force of known masses using torsion balances, as in the Cavendish design, can provide the value of the gravitational constant $G$ [14,15]. Figure 2 summarizes these conclusions.

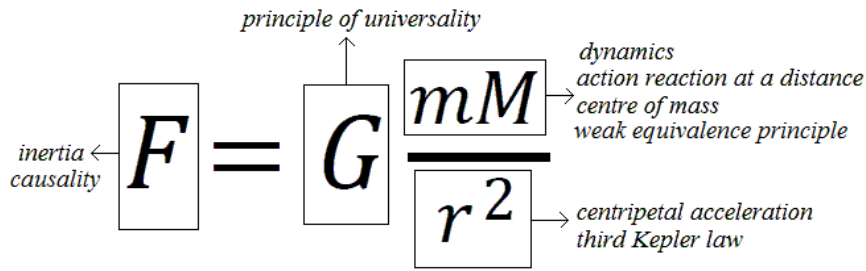

*Figure 2* - The universal law of gravitation along with the elementary concepts and principles behind it.

Einstein´s general theory of relativity uncovers some aspects of gravity in ways unachievable to the Newtonian thinking. For instance, it solves the action at a distance problem. This is done by reinterpreting gravity not in terms of a force, but considering that gravitational mass causes a local spacetime curvature that affects the movement of surrounding objects which, in turn, follow geodesic lines [18]. This corresponds to an application of the 'strong equivalence principle', which says that the Physical laws in an accelerated frame should be no different from the ones experienced in the presence of a gravitational field. Therefore, relativistic effects, such as time dilatation and Lorentz contraction, must be observed in a gravitational field even for slow objects, since gravity can be experienced in velocities much lower than the speed of light. For instance, the free fall of an apple on the Earth´s surface can actually be calculated from the bending of spacetime around Earth, without ever invoking a force law like equation (1) [34,35]. Strictly speaking, for Einstein, gravitational interaction exists through a *spacetime – gravitational mass* coupling so the fundamental concept for Einsten´s theory of gravity is the *gravitational mass* not the *gravitational force*; and if there is no force there is no action at a distance to be considered, as said in the beginning of this paragraph. Newton, however, could not agree with this picture because for him space and time are absolute and independent. Nevertheless, the universality of the gravitational constant, $G$, along with gravitational and inertial masses concepts endure in general relativity. Finally, it should be mentioned that general relativity has an even larger phenomenological range and can be applied to problems Newtonian gravity fails to explain such as the description of the large structure of the Universe [4] and also in the prediction of the recent finding of gravitational waves that has just received the Nobel prize [36]; topics that are out of the scope of this paper.

4. **Conclusions**

The universal law of gravitation represents a great leap of scientific though. By putting together the works of Johannes Kepler and Galileo Galilei, Isaac Newton promoted a profound impact not only in science but in the history of mankind. For instance, Voltaire dedicated part of his works to Newton during the age of Enlightenment [37]. The above presentation brings about one possible and simple route to the law of gravity. Although Newton was inspired by the free fall of an apple in formulating his theory of gravity, the Principia does not mention such fact [2,37]. In this article, the universality of the constant $G$ emerges as a conclusion of a reasoning involving all three Newton´s laws of movement plus the third Kepler law. It also brings $G$, as defined in equation (17), to be essential to recognize the law of gravity as a fundamental and universal postulate. It can be said that it is the law of gravity that makes Newton´s laws of movement, which were developed for Terrestrial objects, valid throughout. Furthermore, it is also important to mention that the universal law of gravitation is related to the action at a distance hypothesis, i.e., that Newton´s third law is valid for non-contact interactions. This characteristic, together with the gravitational mass concept, can be seen as a rudimentary notion of gravitational field hence being important for recognizing equation (1) as fundamental as the other three Newton´s laws of movement and not merely as a corollary of them. On the other hand, the Einsteinian view of gravity substitute the picture of a classical force field by the space-time curvature originated by gravitational mass, thus, overcoming the difficulty of applying Newton´s third law to non contact objects. In the educational aspect, the study of gravity offers a good opportunity to think scientifically, so this contribution intends to bring the students to a moment of creative thinking which goes beyond memorizing a simple list of scientific facts. From this point of view, it is meant to assist the understanding on how rational though, and not data analysis alone, leads to meaningful Physical laws, and ultimately, to the advance of science [38]. Operation of mathematical formulas is required, however at an undergraduate level.

**Acknowledgements**

Many thanks are due to Prof. R. Petrônio for insightful discussions during the GEFONT seminars to acknowledge the 330 years of the Principia.

Dedicated to Miguel and Alina.

**Appendix A** – In the following equations $m_1$ and $m_2$ represent the Earth and the Moon respectively. Correspondingly, one has $\vec{a}_1 = \vec{a}_E^M$ and $\vec{a}_2 = \vec{a}_M^E$. The motion of a system consisting of two interacting bodies of masses $m_1$ and $m_2$ under the influence of external forces, $\vec{F}_1$ and $\vec{F}_2$, can be studied in terms of the following equations:

$$\begin{cases} m_1 \vec{a}_1 = \vec{F}_1 + \vec{f}_{21} \\ m_2 \vec{a}_2 = \vec{F}_2 + \vec{f}_{12} \end{cases} \quad (A1)$$

where $\vec{a}_1$ and $\vec{a}_2$ are the accelerations of each body, $\vec{f}_{12}$ is the force $m_1$ does on $m_2$ and $\vec{f}_{21}$ is the force $m_2$ does on $m_1$. This system can be manipulated to obtain

$$\begin{cases} \vec{A} = \frac{m_1 \vec{a}_1 + m_2 \vec{a}_2}{m_1 + m_2} = \frac{\vec{F}_2}{M} + \frac{\vec{F}_1}{M} + \frac{\vec{f}_{12}}{M} + \frac{\vec{f}_{21}}{M} \\ \vec{a} = \vec{a}_2 - \vec{a}_1 = \frac{\vec{F}_2}{m_2} - \frac{\vec{F}_1}{m_1} + \frac{\vec{f}_{12}}{m_2} - \frac{\vec{f}_{21}}{m_1} \end{cases} \quad (A2)$$

where $\vec{A}$ is the centre of mass acceleration, $\vec{a}$ is the relative acceleration and $M = m_1 + m_2$. Further algebra leads (A2) into equation (7). It is of interest here to consider all the forces being of gravitational nature and, at this point, to make explicit distinction between gravitational and inertial masses. Furthermore, the system in equation (A2) will be analyzed for two bodies moving close enough so the external forces $\vec{F}_1$ and $\vec{F}_2$ are collinear. This is the case of the system treated here since the distance separating the Earth from the Moon is much smaller than their distance to the Sun. Consequently, the gravitational forces of the Sun on the Earth and on the Moon, respectively, are collinear and the corresponding accelerations have nearly the same magnitude too. Accordingly, being $\vec{g}$ the resulting acceleration provided by the Sun, one can write $\vec{F}_1 = m_1^G \vec{g}$, $\vec{F}_2 = m_2^G \vec{g}$ and introducing the inertial masses $m_1^I$ and $m_2^I$ equation (A2) becomes:

$$\begin{cases} \vec{A} = \dfrac{M^G}{M^I}\vec{g} + \dfrac{\vec{f}_{12}+\vec{f}_{21}}{M^I} \\ \vec{a} = \left[\dfrac{m_2^G}{m_2^I} - \dfrac{m_1^G}{m_1^I}\right]\vec{g} + \dfrac{\vec{f}_{12}}{m_2^I} - \dfrac{\vec{f}_{21}}{m_1^I} \end{cases} \quad (A3)$$

where $M^I = m_1^I + m_2^I$ is the total inertial mass of the system and $M^G = m_1^G + m_2^G$ its gravitational counterpart. Equations (A3) are examined in 3 cases below:

**Case i** – $m^I = m^G$, $\vec{f}_{12} + \vec{f}_{21} = 0$ - This is the usual situation where the inertial mass of both bodies are equal to their gravitational mass and the third Newton law holds in its weak form. It follows that:

$$\begin{cases} \vec{A} = \vec{g} \\ \vec{a} = \vec{f}/\mu^I \end{cases} \quad (A4)$$

where $1/\mu^I = 1/m_1^I + 1/m_2^I$ is the reduced inertial mass and $\vec{f} = \vec{f}_{12} = -\vec{f}_{21}$. The centre of mass acceleration is due to the external forces alone while the relative acceleration depends only on the internal force between the two bodies, $\vec{f}$. That means the movement of the system can always be decoupled into the movement of the centre of mass plus the relative movement. Furthermore, if the external forces are null ($\vec{g} = 0$) the linear momentum of the system is conserved since the centre of mass velocity will not vary ($\vec{A} = 0$).

**Case ii** – $m^I = m^G$, $\vec{f}_{12} + \vec{f}_{21} \neq 0$. - This case is examined to investigate what consequences could arise if Newton´s third law, as used in argument **VII**, was not valid. It is admitted that the internal force acts along the line between the centres of the two bodies (strong form of the third law) but may have different magnitudes $\vec{f} = \vec{f}_{12} = -\vec{f}_{21}/\alpha$:

$$\begin{cases} \vec{A} = \vec{g} + \dfrac{(1-\alpha)}{M^I}\vec{f} \\ \vec{a} = \vec{f}/\mu'^I \end{cases} \quad (A5)$$

where $1/\mu'^I = 1/m_2^I + \alpha/m_1^I$ is the reduced inertial mass altered by the factor $\alpha$. The most important feature of equation (A5) is that the linear momentum of the system is not conserved anymore since the centre of mass acceleration is now affected by the internal forces. This would happen regardless of $\vec{g}$ being null and so the system would self-accelerate since the internal forces would control the centre of mass movement. Conservation of linear momentum is lost.

**Case iii** – $m^I \neq m^G$, $\vec{f}_{12} + \vec{f}_{21} = 0$ - This case is examined to investigate what consequences could arise if inertial and gravitational masses would be different but Newton´s third law works at a distance. It follows that:

$$\begin{cases} \vec{A} = \frac{M^G}{M^I}\vec{g} \\ \vec{a} = \left[\frac{m_2^G}{m_2^I} - \frac{m_1^G}{m_1^I}\right]\vec{g} + \frac{\vec{f}}{\mu^I} \end{cases} \quad (A6)$$

If the external forces are null, $\vec{g} = 0$, equation (A6) is reduced to case *i*, $\vec{A} = 0$ and $\vec{a} = \vec{f}/\mu^I$, where the conservation of linear momentum of the system holds and the internal relative movement is decoupled from the centre of mass motion. However, if $\vec{g} \neq 0$, the centre of mass acceleration would not correspond to the acceleration due to gravity. That means an extended body would sense internal stress during a free fall. The gravitational and inertial pulls would be different and, because they are applied to the same body, spurious internal forces would develop causing an eventual self disruption of the falling body. Also, differently from case *i*, the relative motion of the system is now affected by the external forces so it is not possible to decouple the movement.

**Appendix B** – The modern version of Kepler´s problem, which can be found in any advanced mechanics textbook [8, 29], includes the demonstration that the inverse square law of Newtonian gravity is consistent with elliptical orbits through the resolution of non-linear differential equations. This appendix, trails parts of Newton´s original reasoning to explain how to infer the inverse square law using the first and second Kepler´s laws. This is done without resource to differential equations so it can be followed by a first year college student. Moreover, the validity of the Kepler´s third law for elliptical orbits from mechanical energy and angular momentum considerations is shown.

In a letter to the Philosopher John Locke, in 1689, Newton presented a simple proof that the inverse square law can be inferred from an elliptical orbit [29]. If a planet is at the perihelion, P, it will advance to position P' after a small amount of time, which is proportional to the area of the figure PFP' (see figure 2). To treat the planet´s trajectory, Newton adopts the way of thinking used by Galileo in the discussion of the projectile movement using the composition of uniform and accelerated motions [22]. If the time interval is small enough, the segment PQ = $x_P$, will be covered with constant velocity while, at the same time, a uniformly accelerated motion will characterize the segment QP´= $y_P$. Another important finding of Galileo, with direct mentioning by Newton (Principia, Book I, Proposition X, Problem V), was the parabolic path of the projectile. That means that the segments PQ and QP´ are related to each other by $y_P = Cx_P^2$ and also that the segment PP´= $s_P$ will be approximately a parabolic sector. Within the approximation $x_P \sim s_P$ one has that:

$$y_P = Cs_P^2 \tag{B1}$$

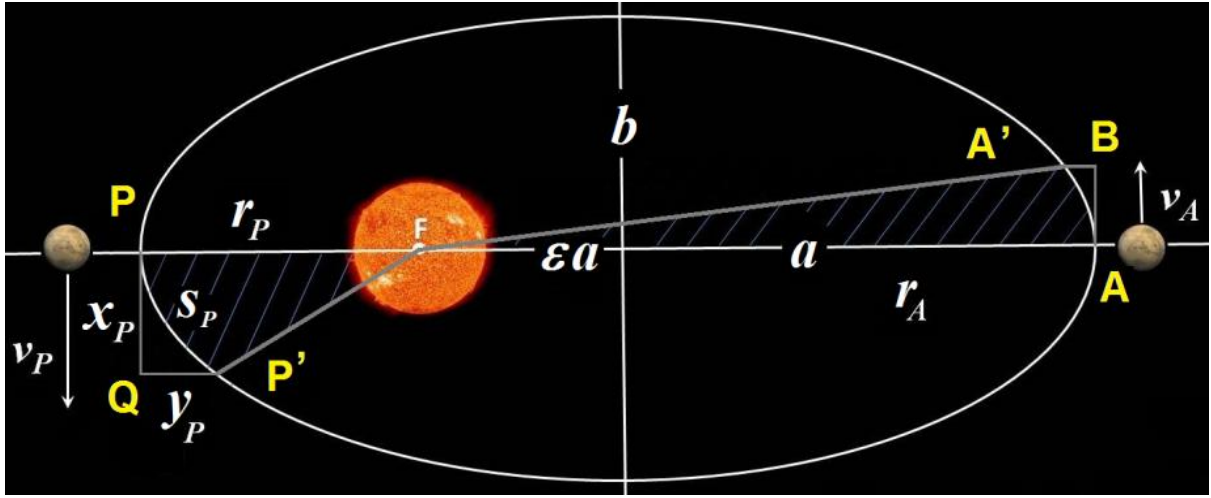

*Figure 2* –*The elliptical orbit of a planet, with eccentricity ε and semi-major axis a, around the Sun located at one focus of the ellipse. The perihelion, at point P, with $r_P = a(1-\varepsilon)$, and the aphelion, at point A, with $r_A = a(1+\varepsilon)$, are taken as reference points. The reflection of the ellipse about the minor axis puts the planet from the perihelion to the aphelion. A second reflection about the major axis turns the velocity upwards reproducing the planet´s revolution. These two operations leave the ellipse unaltered justifying thus equation (B3).*

Because of the symmetry of the ellipse, the shape of the path AA´, that the planet describes when it passes at the aphelion (point A), is equal to the path PP´. That means, if we define the quantities AB = $x_A$, BA´= $y_A$ and AA´= $s_A$, one should have:

$$y_A = Cs_A^2 \tag{B2}$$

analogously to (B1). It follows that

$$y_A/y_P = s_A^2/s_P^2 \tag{B3}$$

since, in view of the symmetry of the ellipse, the constant $C$ is the same in equations (B1) and (B2). Furthermore, if points A and A´ are taken such as the area of AFA´ is the same as PFP´, it follows from Kepler´s second law that the elapsed time the planet takes from A to A´ is the same as the one it takes from P to P´. A second relation involving the parabolic arcs $s_A$ and $s_P$ can then be found applying Kepler´s second law to the areas of PFP´ and AFA´ ($r_P s_P = r_A s_A$) so

$$s_A/s_P = r_P/r_A \tag{B4}$$

If (B4) is used in (B3) one has that

$$y_A/y_P = r_P^2/r_A^2 \tag{B5}$$

Given that the distances the distances $y_A$ and $y_P$ are covered within the same time interval, they are proportional to the corresponding accelerations, corroborating equation (10), and hence to the force. Newton discusses that the parabolic path is an approximation for an ellipse as in the following quote *"... If the ellipsis, by having its centre removed to an infinite distance, degenerates into a parabola, the body will move in this parabola ; and the force, now tending to a centre infinitely remote, will become equable ... which is Galileo´s theorem ..."* [1]. Correspondingly, equation (B5) manifests the connection of the inverse square law with the ellipse. Moreover, it links the projectile motion near the Earth´s surface to the orbital movement of celestial bodies. As mentioned in the above quote, the Galilean parabolic trajectory of a projectile near the Earth´s surface can be proven to be a particular case of a Keplerian orbit [39].

In modern terms, Kepler´s second law is a manifestation of the orbital angular momentum conservation. It states that the area covered by the position vector describing the orbital movement of a planet is proportional to time. In terms of the angular momentum, $L$, of a planet of mass $m$, Kepler´s second law is expressed as [7,8,9]:

$$dA/dt = L/2m \tag{B6}$$

By taking the orbit´s perihelion, P, and aphelion, A, as reference points (see figure 2), the conservation of angular momentum reads

$$mv_P r_P = mv_A r_A \tag{B7}$$

and that permits to write

$$v_A/v_P = [1 - \varepsilon]/[1 + \varepsilon] \tag{B8}$$

where $v_P$ and $v_A$ are the perihelion and aphelion velocities respectively and $\varepsilon$ is the orbit´s eccentricity. In addition, the expression of the mechanical energy for the inverse square force law in equation (1) is given by:

$$E = \frac{1}{2}mv^2 - G\frac{Mm}{r} \tag{B9}$$

Equation (B9) applied to points P and A, can be used together with the conservation of the angular momentum result (B8), to find the velocity at the aphelion:

$$v_A = \sqrt{GM/a}\sqrt{[1-\varepsilon]/[1+\varepsilon]} \tag{B10}$$

From (B10), the angular momentum can be evaluated as

$$L = mv_A r_A = ma\sqrt{GM/a}\sqrt{(1-\varepsilon^2)} \tag{B11}$$

Now, if equation (B6) is integrated over time for the period of an elliptical orbit one has that

$$\pi ab = [L/2m]\,T \tag{B12}$$

where $b$ is semi-minor axis size and $T$ is the period of one revolution which can be isolated in (B12) as:

$$T^2 = 4\pi^2 a^4 (1-\varepsilon^2)\,m^2/L^2 \tag{B13}$$

where $b^2 = a^2(1-\varepsilon^2)$ was employed. Substituting (B11) in (B13) one finds

$$T^2 = [4\pi^2/GM]\,a^3 \tag{B14}$$

which is Kepler´s third law for an elliptical orbit. This result generalizes equation (9) if in equation (B14) the semi-major axis $a$ is identified with the radius of the circular orbit, $r$. It also presents the pre-factor $k$ in equation (9) to be related to the Earth´s mass $M$: $k = 4\pi^2/GM$.